\documentstyle[preprint,aps]{revtex}
\begin{document}
\title{Burst avalanches in solvable models of fibrous materials}
\author{M.\ Kloster, A.\ Hansen and P.\ C.\ Hemmer}
\address{Institutt for fysikk, Norges teknisk-naturvitenskapelige 
universitet, NTNU, N--7034 Trondheim, Norway}
\date{March 15, 1997}
\maketitle
\begin{abstract}
We review 
limiting models for fracture in bundles of fibers, with statistically
distributed thresholds for breakdown of individual fibers. During the
breakdown process, avalanches consisting of simultaneous rupture of
several fibers occur, and the distribution $D(\Delta)$ of the
magnitude $\Delta$ of such avalanches is the central characteristics in
our analysis.  For a bundle of parallel
fibers two limiting models of load sharing
are studied and contrasted: the {\em global} model in which the load
carried by a bursting fiber is equally distributed among the surviving
members, and the {\em local} model in which the nearest surviving 
neighbors take
up the load. For the global model we investigate in particular the
conditions on the threshold distribution which would lead to
anomalous behavior, i.e.\ deviations from the
asymptotics $D(\Delta) \sim \Delta^{-\frac{5}{2}}$, known to be the
generic behavior. For the local
model no universal power-law asymptotics exists, but we show for a
particular threshold distribution how the avalanche distribution
can nevertheless be explicitly calculated in the large-bundle limit.
\end{abstract}
\section{Introduction}
\label{sec1}
When a weak structural element in a material with stochastically
distributed strengths fails, the increased load on the remaining
elements may cause further ruptures, and thus induce a burst
avalanche of a certain size $\Delta$, i.e.\ one in which $\Delta$ elements
fail simultaneously. When the load is further increased, new
avalanches occur. The distribution of avalanche sizes, either at
a fixed load, or the cumulative distribution from zero load until
complete break-down of the material, depends on several factors, in
particular the threshold strength distribution and the mechanism for load 
sharing between the elements.

Due to the complex interplay of failures and redistributions of
local stresses, few analytical results are available in this field; 
computer simulations are commonly applied --- see Herrmann and Roux
\cite{Herrmann} for a review. However, firm analytical results, albeit on 
simplified models, are important in order to develop a deeper understanding 
for universal properties and general trends. In the present article we 
therefore review and study burst events in models of fibrous materials that 
are sufficient simple to allow theoretical treatment.

The models we consider are bundles of $N$ parallel fibers, clamped at
both ends, and stretched by a force $F$ (Fig.\ \ref{fig1}). The individual
fibers in the bundle are assumed to have strength thresholds $f_i$, 
$i=1, 2,\ldots ,N$, which are independent random variables with the same
cumulative distribution function $P(f)$ and corresponding density
function $p(f)$:
\begin{equation}
\mbox{Prob}(f_i < f) = P(f) = \int_0^f \;p(u)\;du.
\label{1}
\end{equation}
Whenever a fiber experiences a force equal to or greater than its strength 
threshold, it breaks immediately and does not contribute to the strength of 
the bundle thereafter.  The models differ, apart from differences in the 
threshold distribution, in how stress is redistributed on the surviving 
fibers when a fiber fails.

A central quantity to be studied in the following is the expected number 
$D(\Delta,N)$ of bursts of size $\Delta$ when the fiber bundle is stretched 
until complete breakdown.

The model of this kind with the longest history \cite{Peirce} is one in which
it is assumed that the fibers obey Hookean elasticity right up to the
breaking point, and that the load distributes itself equally among the
surviving fibers. The model with this democratic load redistribution
is similar to mean-field models in statistical physics, and is called
here the {\em global model}. For large $N$ Daniels \cite{Daniels} was able 
to determine the asymptotic distribution for the bundle strength, a result 
that been refined later \cite{Daniels2}. The distribution of burst 
avalanches was first studied by Hemmer and Hansen \cite{Hemmer}. Their main 
result was that for a large class of threshold distributions $P(f)$ the 
bursts were distributed according to an asymptotic power law,
\begin{equation}
\lim_{N \rightarrow \infty}\frac{D(\Delta)}{N} \simeq \frac{C}{\Delta^{\xi}},
\label{2}
\end{equation}
with a universal exponent
\begin{equation}
\xi = {\textstyle \frac{5}{2}}.
\label{3}
\end{equation}
In Sec.\ \ref{sec2} we show that for special threshold distributions the
power law (\ref{2}) is not obeyed.  

The assumption of global load-sharing among surviving fibers is often 
unrealistic, and it is natural to consider models in which the extra 
stresses by a fiber rupture are taken up by the fibers in the immediate 
vicinity. The extreme version is to assume that only the 
{\em nearest-neighbor} surviving fibers take part in the load-sharing. In a 
one-dimensional geometry, as in Fig.\ \ref{fig1},
presicely two fibers, one on each side, share the extra stress. When
the strength thresholds take only two values, the bundle strength distribution
has been found analytically \cite{Duxbury}. One interesting result is that 
the average bundle strength has a logarithmic size effect. The distribution
of burst avalanches for such models with local load-sharing has not yet been
determined, but simulations \cite{Hansen,Ding} show
that this model is {\em not} in the same universality class as the
global model.  The challenge to determine the burst distribution by
other means than simulations remains, and that this is possible, at least in
a special case (Sec.\ \ref{sec3}), is one of the main results of the present 
article.
\section{The global model}
\label{sec2}
In the global model the total 
force on a fiber bundle is distributed evenly on the
surviving fibers. With a given load $f$ per fiber, all fibers with threshold
less than $f$ have failed, while fibers with thresholds above $f$ survives.
For large $N$ the average number of fibers with thresholds exceeding the
value $f$ will be $N[1-P(f)]$, corresponding to a average total force
\begin{equation}
\langle F\rangle(f) = N f [1-P(f)].
\label{4}
\end{equation}
Let us for the moment assume that $\langle F\rangle(f)$ has a single
maximum. This maximum corresponds then to the value $f = f_c$ for which
$d\langle F\rangle/df$ vanishes. This gives
\begin{equation}
1-P(f_c)-f_c\;p(f_c)=0.
\label{5}
\end{equation}

In Ref.\ \cite{Hemmer} the burst distribution was derived using the 
fiber elongation
$x$ as the independent variable, under the assumption that Hooke's law holds
up to the threshold for breaking. Here, however, we formulate everything
in terms of the force per fiber, $f$, and simplify the derivation
by using directly the fact that the thresholds in a small interval of $f$ are
Poisson distributed.
\subsection{The burst distribution}
\label{sec2a}
Let us consider a small force-per-fiber interval  $(f,f+df)$
in a range where the average force $\langle F\rangle(f)$ increases with $f$.
For a large number $N$
of fibers the expected number of surviving fibers is $N[1-P(f)]$. And
the thresholds in the interval, of which there are $N p(f) df$,
will be Poisson distributed. When $N$ is arbitrary large, the burst
sizes can be arbitrary large in any finite interval of $f$.

Assume that an infinitesimal increase in the external force results in
a break of a fiber with threshold $f$. Then the load that this  fiber
suffered, will be 
redistributed on the $N[1-P(f)]$ remaining fibers; thus they
experience a load increase
\begin{equation}
\delta f = \frac{f}{N[1-P(f)]}.
\label{A1}
\end{equation}
The {\em average} number of fibers that break 
as a result of this load increase is
\begin{equation}
a=a(f)=Np(f)\cdot\delta f = \frac{fp(f)}{1-P(f)}.
\label{A2}
\end{equation}

For a burst of size $\Delta$  the increase in load per fiber will be
a factor $\Delta$ larger than the quantity (\ref{A1}), and an
average number $a(f)\Delta$ will break. The probability that precisely
$\Delta -1$ fibers break as a consequence of the first failure is given
by a Poisson distribution with this average, i.e.\ it equals
\begin{equation}
     \frac{(a\Delta)^{\Delta-1}}{(\Delta-1)!}\;e^{-a\Delta} .
  \label{A3}
\end{equation}
This is not sufficient, however. We must ensure that the thresholds for
these $\Delta -1$ fibers are not so high that the avalanche stops
underway. This requires that at least $n$ of the thresholds are
in the interval
$(f, f+n\delta f)$, for $1\leq n\leq \Delta -1$. In other words,
if we consider the $\Delta$ intervals
$(f,f+\delta f)$, $(f+\delta f, f+ 2\delta f)$, 
\ldots,  $(f+(\Delta-1)\delta,
f+ \Delta \delta f)$, we must find at most $n-1$ thresholds in the
$n$ last intervals. There is the same a priori probability to find a
threshold in any interval. The solution to this  combinatorial problem
is given in Appendix \ref{appa}. The resulting probability to find all
intermediate thresholds weak enough equals $1/\Delta$. Combining this
with (\ref{A3}), we have for the 
probability $\phi (\Delta, f)$ that the breaking
of the first fiber results in a burst of size $\Delta$:
\begin{equation}
\phi (\Delta, f) = \frac{\Delta^{\Delta-1}}{\Delta!}\; a(f)^{\Delta-1}
e^{-a(f)\Delta}.
\label{A4}
\end{equation}

This gives the probability of a burst of size $\Delta$,
as a consequence of a fiber burst due to an infinitesimal increase
in the external load. However, we still have to ensure that the
burst actually {\em starts} with the fiber in question and is not part
of a larger avalanche starting with another, weaker, fiber.
Let us determine the probability  $P_b(f)$ that this initial condition is
fulfilled.

For that purpose consider the $d-1$ fibers with 
the largest thresholds below $f$.
If there is no strength threshold in the interval $(f-\delta f, f)$, at
most one threshold value in the interval $(f-2\delta f, f)$, ... , at most
$d-1$ values in the interval $(f-d\delta f, f)$, then fiber bundle
can not   at any of these previous
$f$-values withstand the external load that forces the fiber
with threshold $f$ to break.
The probability that there are precisely $h$ fiber thresholds in the
interval $(f-\delta f\;d, f)$ equals
\[ \frac{(ad)^h}{h!}\;e^{-ad}.\]
Dividing the interval into $d$ subintervals each of length $\delta f$, the
probability $p_{h,d}$ that these conditions are fulfilled is exactly given by the
solution of the
combinatorial problem in Appendix A: $\; p_{h,d} = 1-h/d$. Summing over
the possible values of $h$, we obtain the probability that the
avalanche can not have started with the failure of a fiber with any of the
$d$ nearest-neighbor threshold values below $f$:
\begin{equation}
P_b(f|d) =\sum_{h=0}^{d-1} \frac{(ad)^h}{h!}\;e^{-ad} (1-\frac{h}{d}) =
(1-a)e^{-ad}\sum_{h=0}^{d-1} 
\frac{(ad)^h}{h!}\; + \;\frac{(ad)^d}{d!} \;e^{-ad}.
\label{Pb}
\end{equation}
Finally we take the limit $d\rightarrow \infty$, for which the last term
vanishes.
For $a>1$ the sum must vanish since the left-hand side of (\ref{Pb}) is
non-negative, while the factor $(1-a)$ is negative. For $a<1$, on the
other hand, we find
\begin{equation}
P_b(f) = \lim_{d\rightarrow \infty} P_b(f|d) = 1- a,
\end{equation}
where $a = a(f)$. The physical explanation of the different behavior
for $a>1$ and $a \leq 1$ is straightforward: The
maximum of the total force on the bundle occurs at $f_c$ for which $a(f_c)
=1$, see Eqs.\ (\ref{5}) and (\ref{A2}),
so that  $a(f)>1$ corresponds to $f$ values almost certainly involved in
the final catastrophical burst. The region of 
interest for us is therefore when
$a(f) \leq 1$, where avalanches on a microscopic scale occur. This is
accordance with what we found in the beginning of this section, viz.\ that
the burst of a fiber with threshold $f$ leads immediately to a
average number $a(f)$ of additional failures.

Summing up, we obtain the probability that the fiber with threshold $f$
is the first  fiber in an avalanche of size $\Delta$  as the product
\begin{equation}
\Phi(f) = \phi(\Delta, f) 
P_b(f) = \frac{\Delta^{\Delta-1}}{\Delta !}\; a(f)^{\Delta -1}
e^{-a(f)\Delta} [1-a(f)],
\label{prob}
\end{equation}
where $a(f)$ is given by Eq.\ (\ref{A2}),
\[ a(f) = \frac{f\;p(f)}{1-P(f)}.\]

Since the number of fibers with 
threshold in $(f,f+\delta f)$ is $N\;p(f)\;df$,
the burst distribution is given by
\begin{equation}
\frac{D(\Delta)}{N} = \frac{1}{N}\int_0^{f_c} \Phi (f) p(f)\;df =
\frac{\Delta^{\Delta -1}}{\Delta !}
\int_0^{f_c} a(f)^{\Delta -1}e^{-a(f)\Delta}
\left[1-a(f)\right] \;p(f)\;df.
\label{Delta}
\end{equation}

For large $\Delta$ the maximum contribution to the integral comes
from the neighborhood of the upper integration limit, since $a(f)\;e^{-a(f)}$
is maximal for $a(f)=1$, i.e.\ for 
$f=f_c$. Expansion around the saddle point and
integration yields the asymptotic behavior
\begin{equation}
D(\Delta)/N \propto \Delta^{-\frac{5}{2}},
\label{asymp}
\end{equation}
universal for those threshold distributions for which the assumption of
a single maximum of $\langle F\rangle(f)$ is valid.

Note that if the experiment had been stopped before complete breakdown,
at a force per fiber
$f_m < f_c$, the asymptotic behavior would have been {\em exponential}
rather than a power law:
\begin{equation}
D(\Delta)/N \propto \Delta^{-\frac{5}{2}} e^{-[a(f_m)-1-\ln a(f_m)]\Delta}.
\end{equation}
In the form
\begin{equation}
D(\Delta) \propto \Delta^{-\eta}
e^{-\Delta/\Delta_0},\hspace{5mm}\mbox{with}\hspace{5mm} \Delta_0 \propto
(f_c-f)^{-\nu},
\label{crit}
\end{equation}
the breakdown
process is similar to a critical phenomena with a critical point at
total breakdown  \cite{Hansen2}. 
The distribution follows a power law with index $\eta =
\frac{5}{2}$ with
a cutoff that diverges at total failure with an index $\nu = \frac{1}{2}$.

What happens when the average strength $\langle F\rangle(f)$ curve does not have
a unique maximum? If it has several parabolic maxima, and the absolute
maximum does not come first (i.e.\ at the lowest $f$ value),
then there will be several avalanches of macroscopic size in the sense that
a finite fraction of the $N$ fibers break simultaneously \cite{Lee}.
The asymptotics (\ref{asymp}) 
is thereby unaffected, however.   We turn next to
threshold distributions that are more interesting because they
lead to different asymptotics.
\subsection{Strong threshold distributions}
\label{sec2b}
Rather than consider bundle strength functions $\langle F\rangle(f)$ with
several parabolic maxima, we study now cases in which there is no
such maximum.
We are  particularly interested in the asymptotics of the burst
distributions.

Model examples of such threshold distributions are
\begin{equation}
P(f)=\left\{ \begin{array}{ll}
0 & \mbox{for }\; f \leq f_0 \\
1-[1+(f-f_0)/f_r]^{-\alpha} & \mbox{for }\; f > f_0
\end{array} \right.
\label{brede}
\end{equation}
Here $\alpha$ and $f_0$ are positive parameters, and $f_r$ is a reference
quantity which we for simplicity put equal to unity in the following.
This class of threshold distributions is rich enough to exhibit several
qualitatively different avalanche distributions.

The corresponding macroscopic bundle strength per fiber is, according to
Eq.\ (\ref{4}),
\begin{equation}
\frac{\langle F\rangle(f)}{N} = \left\{\begin{array}{ll}
f & \mbox{for}\;\;\;f\leq f_0  \\
\frac{f}{(1+f-f_0)^{\alpha}} &\mbox{for} \;\;\; f>f_0
\end{array} \right.
\label{macro}
\end{equation}
In Fig.\ \ref{fig2} 
some threshold distributions 
$p(f)$ and the corresponding macroscopic force curves
$\langle F\rangle (f)$ are sketched.

The distribution of avalanche sizes is given by Eq.\ (\ref{Delta}).
In the present case the function $a(f)$ takes the form
\begin{equation}
a(f) = \frac{fp(f)}{1-P(x)} = \frac{\alpha f}{1+f-f_0}.
\label{af}
\end{equation}
A simple special case is $f_0=1$, corresponding to
\[ p(f) =
\alpha \; f^{-\alpha -1} \hspace{1cm} \mbox{for} \;\; f \geq 1,\]
 since then the function (\ref{af}) is independent of $f$:
\[ a(f) = \alpha.\]
This gives at once
\begin{equation}
\frac{D(\Delta)}{N} =\frac{1-\alpha}{\alpha} \;
\frac{\Delta^{\Delta - 1}}{\Delta !} \left[\alpha e^{-\alpha}\right]^{\Delta}
\simeq \frac{1-\alpha}{\alpha\sqrt{2\pi}}\;\Delta^{-\frac{3}{2}}
\left[\alpha e^{1-\alpha}\right]^{\Delta}.
\label{one}
\end{equation}

In other cases it is advantageous to change 
integration variable in Eq.\ (\ref{Delta})
from $f$ to $a$:
\begin{equation}
\frac{D(\Delta)}{N} = \frac{\Delta^{\Delta -1}}{e^{\Delta}\Delta !}\;
\frac{1}{\alpha^{\alpha -1}(1-f_0)^{\alpha}} 
\int\limits_{\alpha f_0}^{\alpha}
(\alpha - a)^{\alpha -1}(1-a)a^{-1}\left(ae^{1-a}\right)^{\Delta}\;da.
\label{integral}
\end{equation}
The asymptotics for large $\Delta$,
beyond the  $\Delta^{-\frac{3}{2}}$
dependence of the prefactor, is determined by the $\Delta$-dependent
factor in the integrand.  The maximum of $ae^{1-a}$ is unity, obtained for
$a = 1$, and the
asymptotics depends crucially on whether $a=1$ falls outside the range
of integration, or inside
(including the border). If the maximum falls
inside the range of integration the
$D(\Delta) \propto \Delta^{-\frac{5}{2}}$ dependence
remains. A special case of this is $\alpha = 1$, for which  the
maximum of the integrand is located at the integration limit and the
macroscopic force has a ``quadratic'' maximum {\em at infinity}.

Another special case is  $\alpha f_0=1$ (and $\alpha <1$), for which
again the standard asymptotics $\Delta^{-\frac{5}{2}}$ is valid. In this
instance the macroscopic force has a quadratic {\em minimum} at
$f=f_0$ (see Fig.\ \ref{fig2} 
for $\alpha=\frac{1}{2}$), and  critical behavior arises 
just as well from a minimum as from a maximum.

In the remaining cases, in which $a=1$ is not within the range of
integration in Eq.\ (\ref{integral}), the avalanche distribution is always a
power law with an exponential cut-off,
\begin{equation}
\frac{D(\Delta)}{N} \simeq \Delta^{-\xi} \; A^{\Delta}.
\end{equation}
Here $\xi$ and $A$ depend 
on the parameter values $f_0$ and $\alpha$, however.
This is easy to understand. Since
\begin{equation}
\frac{da(f)}{df} = \frac{\alpha (1-f_0)}{(1+f-f_0)^2},
\end{equation}
we see that $a(f)$ is a monotonically
decreasing function for $f_0>1$, so that
the maximum of $ae^{1-a}$ is obtained at the lower limit $f=f_0$, where
$a=\alpha f_0$. The asymptotics
\begin{equation}
D(\Delta) \propto \Delta^{-\frac{5}{2}} \left(\alpha f_0e^{1-\alpha f_0}
\right)^{\Delta}
\label{two}
\end{equation}
follows.

This is true merely for $\alpha f_0 < 1$, however. For $\alpha f_0 > 1$
the macroscopic force $\langle F\rangle(f)$ {\em decreases} near $f=f_0$ so that
a macroscopic burst takes place at a force $f_0$ per fiber, and stabilization
is obtained
at a larger force $f_1$ (Fig.\ \ref{fig2}). 
The subsequent bursts have an asymptotics
\begin{equation}
D(\Delta) \propto \Delta^{-\frac{5}{2}} \left(a(f_1) e^{1-a(f_1)}
\right)^{\Delta},
\end{equation}
determined by the neighborhood of $f=f_1$.

For $f_0<1$, the maximum of $ae^{1-a}$ is obtained at $f=\infty$, leading
to the asymptotics
\begin{equation}
D(\Delta) \propto \Delta^{-\frac{3}{2}-\alpha} \left(\alpha e^{1-\alpha}
\right)^{\Delta},
\end{equation}
reflecting the power-law behavior of the integrand at infinity.

The results are summarized in Table \ref{tab1}.
Note that the $f_0=1$ result (\ref{one})  cannot be obtained by
putting $f_0=1$ in Eq.\ (\ref{two})
since in (\ref{integral}) the order of the limits $\Delta \rightarrow \infty$ and $f_0 \rightarrow
1$ is crucial.
\section{The local model}
\label{sec3}
The assumption of global loadsharing among 
surviving fibers is often unrealistic,
since fibers in the neighborhood of 
the failed fiber are expected to take most
of the load increase. The extreme form for local load redistribution is that
all extra stresses caused by a fiber 
failure are taken up by the {\em nearest-neighbor}
surviving fibers.

The simplest geometry is one-dimensional so that the $N$ fibers are
ordered linearly, without or with periodic boundary conditions 
(Fig.\ \ref{fig1}). In
this case precisely two fibers, one on each side, take up, and divide
equally, the extra stress. At a total force $F_{tot}$
on the bundle the force on a fiber  surrounded by $n_l$
previously failed fibers on the left-hand side, and $n_r$ on the right-hand
side,  is then
\begin{equation}
\frac{F_{tot}}{N} 
\left(1+{\textstyle \frac{1}{2}}(n_l+n_r)\right) = x(2+n_l+n_r).
\label{HP}
\end{equation}
Here
\begin{equation}
x=\frac{F_{tot}}{2N},
\label{x}
\end{equation}
one-half the force-per-fiber, is a convenient variable to use as the
driving {\em force parameter}.
This model has been discussed previously \cite{Harlow} 
for a different purpose.
Preliminary studies \cite{Hansen,Ding} of the
avalanche distribution for some threshold strength distributions have not
yielded analytical results but  simulation  results that show convincingly
that the local model is {\em not} in
the same univerality class as the global model.

In order to obtain explicit results we assume for the fiber strengths
the simplest possible case, a {\em uniform} 
threshold distribution. In units of the maximum threshold:
\begin{equation}
P(f) = \left \{ \begin{array}{ll}
f& \hspace{3mm} \mbox{for } 0 \leq f < 1 \\
1 & \hspace{3mm} \mbox{for } f \geq 1.
\end{array} \right.
\end{equation}

Avalanches in the local and the global models have  different characters.
In the local model an avalanche unroll with one failure
acting as the seed. If many neighboring fibers have failed, the load on the
fibers on each side is high, and if they burst the load on the new neighbors
will be even higher, etc. In this way a weak region in the bundle may be
responsible for the failure of the whole bundle. For a large number $N$ of
fibers the probability of a weak region somewhere is high, and this
explains in a qualitative way that 
the maximum load the bundle are able to carry
does not increase proportional to $N$, but slower than linear.

The load distribution rule (\ref{HP}) implies that an avalanche of size
$\Delta$  does necessarily
lead to a complete breakdown of the whole bundle if the external force
is too high, i.e., if $x$ exceeds a critical value $ x_{max}$.
Since here a fiber can at most take a load
of unity, we have
\begin{equation} x_{max} = \frac{1}{\Delta + 2}.
\label{max}
\end{equation}

The strategy of the derivation is to first establish a set of recursion
relations between  quantities that give probabilities
of certain configurations at fixed external 
force, i.e., at fixed $x$. Afterwards
(Sec.\ \ref{sec3b})
we connect this with the size distribution of avalanches for all $x$ up to
the critical value $x_{max}$.
\subsection{Recursion relations}
\label{sec3a}
We will use the terminology that the
{\em magnitude} of an avalanche is 
the number of failing fibers in the avalanche, and
the {\em length} of an avalanche is the number of fibers between the
nearest surviving fibers on each side of the avalanche. The length can
be larger than the magnitude since it may include fibers that have failed in
previous avalanches.

We define $S(l;x)$, the {\em gap probability}, to 
be the probability (at given force parameter $x$) that
in a selected
region of $l$ consecutive fibers all fibers have failed, assuming
the two fibers on each side to be intact. We
let $S(0;x) = 1$ by definition.

Another central quantity is the probability density $p(l,a;x)$. We define
it by selecting a  region of $l$ consequtive fibers, and let
 $p(l,a;x)\;dx$ be the probability that a force increase from
$x$ to $x+dx$ leads to an avalanche of this length $l$ and of magnitude $a$.

The state at force parameter $x$ that 
all $L$ fibers have failed  must have appeared
for some force parameter in the 
range $(0,x)$, and by a burst of some magnitude $a$ in the
range $(1,L)$. Thus
\begin{equation}
S(L;x) = \sum_{a=1}^L \int_0^x p(L,a;y)\;dy.
\label{rec1}
\end{equation}

Let us  now obtain expressions for the probability density
$p(L,a;x)$, first for the special case that the  magnitude $a$ is unity.
Just one fiber fails in this burst, and in
an avalanche of length $L$ therefore $L-1$ of the neighboring fibers must
already have failed. By (\ref{HP}) the force on the fiber just before it
fails is $(L+1)x$, and for the uniform distribution  the
probability that it fails due to a force parameter increase  $dx$ is just
$(L+1)dx$.
The probability of the burst of magnitude $1$ to occur when $x\rightarrow x+
dx$ is this probability of failure of the single remaining fiber, $(L+1)dx$,
times the probability that the $L-1$ 
neighbors have already failed. The latter
is given by the appropriate gap probabilities.
Since the position of the failing fiber is arbitrary
in the interval we have
\begin{equation}
p(L,1;x)\;dx = \sum_{i=0}^{L-1} S(L-i-1;x)\; (L+1)dx\; S(i;x).
\label{rec2}
\end{equation}

We next consider expressions for the probability $p(L,a;x)$ with an internal
avalanche of magnitude $a$ larger than unity. For that purpose
we introduce two new quantities: Let $p_l(L+1,a;x)\;dx$ be the
probability that a fiber fails because a 
force parameter increase $x\rightarrow x+dx$
starts, on its right-hand side, an avalanche of magnitude $a$
(not counting the ultimate fiber on the left-hand side)
and of length $L$.
Similarly $p_r(L+1,a;x)\;dx $ is the probability that a fiber fails because
the force parameter increase $x\rightarrow x+dx$ 
starts, on its left-hand side, an avalanche of magnitude $a$ and of length 
$L$.

Consider the event described by $p(L,a;x)$, and let the last of the $a$
fibers that fail be fiber ${\cal F}$. 
The force  distribution mechanism in the
local model implies that ${\cal F}$ is either the leftmost or the
rightmost   of the $a$ fibers. The first possibility implies that
the increase $x\rightarrow x+dx$ induces
the first failure to the right of ${\cal F}$, which starts an avalanche
of magnitude $a-1$ and length $i$, say, to the right of ${\cal F}$. Here
$a-1 \leq i \leq L-1$, of course, 
and ${\cal F}$ must have $L-i-1$ previously
failed fibers on its left-hand side.

Including all possibilities we have
\begin{eqnarray}
p(L,a;x) =\sum_{i=a-1}^{L-1} \left[ S(L-i-1;x)\;p_l(i+1,a-1;x) +\right.
\nonumber\\
 \left.   p_r(i+1,a-1;x)\;S(L-i-1;x)\right],
\label{rec3}
\end{eqnarray}
where the second term represents events in which the first failure occurs
to the {\em left} of ${\cal F}$.

On the other hand we want to express 
$p_l(l+1,a;x)$ and $p_r(l+1,a;x)$ in terms of previously
defined quantities. For magnitude $a=1$ this 
is relatively simple. Let the single
fiber (call it ${\cal G}$) that starts 
the process have $n$ already failed fibers to the right and
$l-n-1$ fibers to the left, with $0\leq n \leq l-1$.  The probability that
fiber ${\cal G}$ fails under the load increase  $x\rightarrow x+dx$ is
$(l+1)dx$ for the uniform threshold
distribution.  Since the failure
of fiber ${\cal G}$ causes a load increase $(n+1)x$ on the left fiber,
the probability of its failure is  $(n+1)x$.
We have,  when all possible positions of ${\cal G}$ are taken into account,
\begin{equation}
p_l(l+1,1;x)\;dx = \sum_{n=0}^{l-1} S(l-n-1;x)\; (l+1)dx\; S(n;x) (n+1)x
\label{rec4}
\end{equation}
Similarly,
\begin{equation}
p_r(l+1,1;x) =\sum_{n=0}^{l-1} S(l-n-1;x)(l+1)S(n;x)(l-n)x = p_l(l+1,1;x).
\label{rec5}
\end{equation}

The corresponding expressions for $p_r(L+1,a;x)$ and $p_l(L+1,a;x)$, with
$a$ larger than unity, are
more complicated. The internal avalanche started by $x\rightarrow x+dx$
proceeds so that the final failure is either the leftmost or the
rightmost of the $a$ fibers, or both. If both go simultaneously, we make the
arbitrary definition that in such a case the right-hand neighbor fails first.
This secures a unique sequential ordering of failures.

Consider first $p_r(L+1,a+1;x)$, 
and denote the last surviving fiber on the right-hand
side as ${\cal F}$. Assume first that the rightmost of the $a$
internal fibers fails last, and let this fiber have $i$ fibers on its
left-hand side and $L-i-1$ failed fibers on the right-hand side. The
probability that this right-hand side fiber fails under $x\rightarrow
x+dx$ with an internal avalanche of magnitude $a$ and length $i$ is
just $p_r(i+1,a;x)\;dx$, and the probability of finding $L-i-1$ failed
fibers on the right-hand side is given by the gap probability $S(L-i-1;x)$.
After the rightmost internal fiber has failed the load increase on the
${\cal F}$ is $(i+1)x$, which also equals its probability of
failure. The other alternative is that the leftmost of
the $a$ internal fibers fails last, with, say,
$i$ fibers on its right-hand side, and $L-i-1$ failed fibers on its left.
Then the extra load increase on ${\cal F}$, and hence its probability of
failure, is $(L-i)x$. Including all possible positions $i$ we
end up with
\begin{eqnarray}
p_r(L+1,a+1;x) =\sum_{i=0}^{L-1} S(L-i-1;x)\left[p_r(i+1,a;x)(i+1)x+\right.
\nonumber\\
\left.p_l(i+1,a;x)(L-i)x\right].
\label{rec6}
\end{eqnarray}

By a similar argument the corresponding expression for
$p_l(L+1,a+1;x)$ is built up. However, 
when in this case the rightmost of the
internal fibers fails last we must add the probability
that the leftmost and the rightmost fibers are simultaneously overburdened.
Letting $p_2(i+2,a;x)\;dx$ be the 
probability that an avalanche of length $i$
and size $a$ makes {\em both} neighbor fibers fail, we have
\begin{eqnarray}
p_l(L+1,a+1;x)=\sum_{i=0}^{L-1}S(L-i-1;x)\left[p_l(i+1,a;x)(i+1)x+\right.
\nonumber\\
\left.p_r(i+1,a;x)(L-i)x+p_2(i+2,a;x)\right].
\label{rec7}
\end{eqnarray}

Finally, the recursion relations for $p_2$ close 
the set. One sees easily that
when the failure of the two end fibers is caused by a single internal
fiber burst due to the force increase from $x$ to $x+dx$, we have
\begin{equation}
p_2(L+2,1;x)=\sum_{i=0}^{L-1}S(L-i-1;x)(L+1)S(i;x)(i+1)x(L-i)x.
\label{rec8}
\end{equation}
Here $(L+1)dx$ is the probability that the single fiber fails, $(i+1)x$
is the probability that the $(i+1)$ new failures on the right makes the
left-hand-side fiber break, while $(L-i)x$ is the probability that the
$(L-i)$ new failures on the left makes the right-hand-side fiber break.

When the failure of the two end fibers are caused by an internal
avalanche involving $a>1$ fibers we may argue along the same lines as for
$p_l$, with the result
\begin{eqnarray}
p_2(L+2,a+1;x)=
\sum_{i=0}^{L-1}S(L-i-1;x)\left[p_l(i+1,a;x)(i+1)x(L-i)x       \right.
\nonumber\\
\left.+\left(p_r(i+1,a;x)(L-i)x+p_2(i+2,a;x)\right)(i+1)x\right].
\end{eqnarray}

We can simplify the set of equations somewhat by introducing the sum
$p_s = p_l+p_r$, with the result
\begin{eqnarray}
S(L;x) &=& \sum_{a=1}^L \int_0^xp(L,a;y)\;dy  \label{l1}\\
p(L,1;x) &=& \sum_{i=0}^{L-1} S(L-i-1;x)(L+1)S(i;x) \label{l2} \\
p(L,a+1;x) &=& \sum_{i=0}^{L-1} S(L-i-1;x) p_s(i+1,a;x)\label{l3}\\
p_s(L+1,1;x)  &=& \sum_{i=0}^{L-1} S(L-i-1;x)(L+1)S(i;x)(L+1)x \label{l4}\\
p_s(L+1,a+1;x)   &=& \sum_{i=0}^{L-1}
S(L-i-1;x)\left[p_s(i+1,a;x)\right. \nonumber \\
& & \left.+p_2(i+2,a;x)\right]\label{l5}\\
p_2(L+2,1;x)   &=& \sum_{i=0}^{L-1} S(L-i-1;x)(L+1)S(i;x)
\nonumber\\
& & (i+1)x(L-i)x \label{l6}                  \\
p_2(L+2,a+1;x)   
&=& \sum_{i=0}^{L-1} S(L-i-1;x)(i+1)x\left[p_s(i+1,a;x)(L-i)x \right.
\nonumber \\
& &\left.+p_2(i+2,a;x)\right] \label{17}  
\end{eqnarray}

Starting with $S(0;x)=1$, one easily proves by induction the following
$x$-dependence of all quantities involved:
\begin{eqnarray}
S(L;x) &=& S(L)x^L \label{m1}\\
p(L,a;x) &=& p(L,a)x^{L-1} \label{m2}\\
p_s(L+1,a;x) &=& p_s(L+1,a)x^L \label{m3}\\
p_{\Delta}(L+2,a;x) &=& p_{\Delta}(L+2,a)x^{L+1} \label{m4}
\end{eqnarray}

In $x$-independent form the recursion relations then 
take the form (with $a>0$)
\begin{eqnarray}
S(L)& =& L^{-1}\sum_{a=1}^L p(L,a)  \label{n1}\\
p(L,1)& =& \sum_{i=0}^{L-1} S(L-i-1)(L+1)S(i) \label{n2}\\
p(L,a+1)& =& \sum_{i=0}^{L-1} S(L-i-1) p_s(i+1,a)\label{n3}     \\
p_s(L+1,1)&  =& \sum_{i=0}^{L-1} S(L-i-1)(L+1)S(i)(L+1) \label{n4}\\
p_s(L+1,a+1)& 
=& \sum_{i=0}^{L-1} S(L-i-1)\left[p_s(i+1,a)+p_2(i+2,a)\right]\label{n5}\\
p_2(L+2,1)&   =& \sum_{i=0}^{L-1} S(L-i-1)(L+1)S(i)(i+1)(L-i) \label{n6}\\
p_2(L+2,a+1) &  
=& \sum_{i=0}^{L-1} S(L-i-1)(i+1)\left[p_s(i+1,a)(L-i)+\right.
\nonumber \\
& &\left. p_2(i+2,a)\right] \label{n7}
\end{eqnarray}

Let us finally note that the feature of the uniform distribution
that makes the derivation simpler than for other distributions is that
the probability for failure of a fiber is given by the load increase,
{\em independent} of the actual load level.

We can now calculate recursively $S(L;x)$ and $p(L,a;x)$ for integer
$L$ and $a$. By (\ref{m1})--(\ref{m4}) the $x$ dependence is trivial.
\subsection{The asymptotic burst distribution}
\label{sec3b}
In order to use the quantitative information obtained above we must first
determine  the survival probability $P_s(N,x)$
that a fiber bundle is able to tolerate a force per fiber equal to $2x$.
Noting that in this model avalanches are local phenomena, and that two
 failed fibers are only correlated
when all fibers inbetween have failed, the survival probability $P_s(N,x)$
is expected to depend {\em exponentially} on 
the length $N$ for large $N$, so that
\begin{equation}
\lim_{N\rightarrow \infty} N^{-1} \ln P_s(N,x) = t(x)
\label{survival}
\end{equation}
is finite.  The exponential form of the survival probability is discussed
and confirmed in other studies  \cite{Harlow2,Kuo,Zhang}.

We assume periodic boundary conditions, and 
number the fibers from an arbitrary starting point.
We define $P_f(n,L;x)$ to be the probability, at force parameter $x$, that
among the $n$ first fibers there is no fatal burst, and that the last $L$
fibers of these have all failed. Fiber 
number $n+1$ is assumed to hold. We will now
establish a recursion relation between the $P_f(n,L;x)$.

Consider a region of $n+1+L$ fibers in which no fatal burst has occurred,
and where the last $L$ fibers have failed. The probability of this
configuration is $P_f(n+1+L,L;x)$. In the region to the left of
fiber number $n+1$
let the length of the region of broken fibers that contain
fiber number $n$ be $i$, where  $i$ may take all values between zero
(if fiber number $n$ is intact) and $M(x) =[x^{-1}-2]$.
The region to the right of fiber number $n+1$ has $L$ broken fibers, and the
probability of this is $S(L;x)$.
This gives the recursion relation
\begin{equation}
P_f(n+1+L,L;x) = \sum_{i=0}^{M(x)-L} P_f(n,i;x)[1-(i+L+2)x]S(L;x).
\label{Pf}
\end{equation}
The last factor is the probability that fiber number $n+1$, which has
$i+L$ failed neighbors, holds.

Insertion of the product form $P_f(n,L;x) \simeq t(x)^n P_f(L;x)$ into
(\ref{Pf}) yields the following equations for the $P_f(i;x)$:
\begin{equation}
P_f(L;x)-\sum_{i=0}^{M(x)} [1-(i+L+2)x]S(L;x)t(x)^{-L-1}\;P_f(i;x).
\label{pf}
\end{equation}
It is consistent to let $P_f(0;x)=1$.
Since $L$ may take the values $0,1,\ldots ,M(x)$, (\ref{pf}) 
is a set of $M(x)+1$
homogeneous equations for the $M(x)+1$ quantities $P_f(i;x)$. The system
determinant of the equation set must vanish, and this determines $t(x)$ for
a given force parameter $x$. With $P_f(0;x)=1$ all quantities can then
determined. The practical solution procedure is by iteration.

From the definitions of $P_f(n,L;x)$ and $S(L;x)$ it follows that the ratio
\[ \frac{P_f(n,L;x)}{S(L;x)}\]
is the probabilility, at force parameter 
$x$, that  among the first $n$ fibers
there is no fatal burst, {\em given} that 
there are $L_1$ failed fibers on the
right-hand side. Then
\begin{equation}
\frac{P_f(n,L_1;x)}{S(L;x)} \;p(L_1,\Delta;x)\; dx
\label{pro}
\end{equation}
is the probability that an increase of the 
force parameter from $x$ to $x+dx$
starts an avalanche of size $\Delta$ 
and length $L$, so that afterwards there
is no fatal burst among the $n$ fibers on the left-hand side.

Finally we want to determine the probability 
for a burst of size $\Delta$ in
a system of $N$ fibers in a ring configuration (Fig.\ \ref{fig1}). 
On the left of a
selected fiber $\bf f$ we consider a region of $n$ fibers, and on the
right a region of $N-n-1$ fibers.
The probability that the force parameter increase $x\rightarrow x+dx$
induces a burst of size $\Delta$ and length $L_1$ to the left of
$\bf f$ that holds is given by (\ref{pro}). Here $\Delta < L_1 \leq n$, of
course. On the right-hand side
of $\bf f$ a number $L_2$ fibers adjacent to $\bf f$ may have failed. (Here
$L_2$ is less than the remaining number
$N-n-1$ of fibers.) The probability of such a configuration (with no fatal
burst) is $P_f(N-n-1,L_2;x)$. We must also take into account that the
fiber $\bf f$ itself, with $L_1+L_2$ failed neighboring fibers, must hold,
the probability of which is $[1-(L_1+L_2+2)x]$.

When we take this together, sum over the possible values of $L_1$, $L_2$ and
$n$, and integrate over $x$, we obtain
\begin{eqnarray}
D(\Delta) = 
\int\limits_0^{1/(\Delta +2)} \sum_{n=1}^N \sum_{L_1=\Delta}^{M(x)}
\sum_{L_2=0}^{M(x)-L_1} \frac{P_f(n,L_1;x)}{S(L;x)}p(L_1,\Delta;x)
\nonumber\\
P_f(N-n-1,L_2;x)[1-(L_1+L_2+2)x]\;dx.
\end{eqnarray}
Using the product property $P_f(n,L;x) \simeq t(x)^N P_f(L;x)$ the sum over
$n$ simply yields a factor $N$, and we find
\begin{eqnarray}
&\frac{D(\Delta)}{N} =
\int\limits_0^{1/(\Delta+2)}\sum_{L_1=0}^{M(x)}\sum_{L_2=0}^{M(x)-L_1}
\frac{P_f(L_1;x)}{S(L_1;x)}p(L_1,\Delta;x)P_f(L_2;x)t(x)^{N-1}\nonumber\\
&[1-(L_1+L_2+2)x]dx.
\label{864}
\end{eqnarray}

This may now be evaluated. The results for a bundle of $N=20\; 000$ are 
shown in Table \ref{tab2}, together with simulation results for 4 000 000 
bundles, each having 20 000 fibers.

The agreement between the simulation data and the theoretical data is,
as we see, extremely satisfactory.

An analysis of the burst distribution obtained for this local model
shows that the distribution does not follow a power law except for
small values of $\Delta$ (Fig.\ \ref{fig3}).
If one nevertheless does a linear regression analysis on this part of the 
data set, the effective power would be of the order $5$, considerable 
larger than the ``mean-field'' value $\frac{5}{2}$ for the global model 
\cite{Hansen,Ding}.
\subsection{Size-dependent bundle strength}
\label{sec3c}
Let us  now attempt to find an simple 
estimate for the maximal force per fiber
that the fiber bundle can tolerate. In order to do that we assume that the
fatal burst occurs in a region where no fibers have previously failed
so that the burst has
the same magnitude and length. We know that a single burst of length
$\Delta = x^{-1}-2$ is fatal, Eq.\ (\ref{max}), so our criterion is
simply
\begin{equation}
D(x^{-1}-2) = 1.
\label{fatal}
\end{equation}

If we take into account that the two fibers adjacent to the burst should
hold, and ignore the rest of the bundle, the gap distribution would be
\begin{equation}
N^{-1} D(\Delta) \approx 
\int\limits_0^{1/(\Delta+2)}[1-(2+\Delta)x]^2p(\Delta,
\Delta;x)\;dx 
=\frac{2p(\Delta,\Delta)}{\Delta(\Delta+1)(\Delta+2)^{\Delta+1}}.
\end{equation}
With the abbreviation
\[   R_{\Delta} = \frac{p(\Delta,\Delta)}{(\Delta -1) !},\]
we have
\begin{equation}
D(\Delta)/N \approx \frac{2(\Delta+2)!}{\Delta^2(\Delta+1)^2
(\Delta+2)^{\Delta+2}}
R_{\Delta}
\simeq \frac{\sqrt{8\pi(\Delta +2)}\;}{\Delta^2(\Delta+1)^2}e^{-\Delta-2}
\;R_{\Delta},
\end{equation}
using Stirling's formula.

Taking logarithms we have
\begin{equation}
\ln D(\Delta)-\ln N = -
(\Delta + 2)\left[1+ \frac{\ln R_{\Delta}}{\Delta+2} + {\cal O}
\left(\frac{\ln \Delta}{\Delta}\right)
\right] \simeq - (\Delta +2),
\end{equation}
using the result (\ref{conv}) of Appendix \ref{appb}
for $R_{\Delta}$ when $\Delta$ is large.

The failure criterion (\ref{fatal}) then takes the form
\[    \ln N \simeq \frac{1}{x}.\]
Since $x=F/2N$ we have the following estimate for the maximum force $F$
that the fiber bundle
can tolerate before complete failure:
\begin{equation}
F \simeq \frac{2N}{\ln  N}.
\label{Ffatal}
\end{equation}
Thus the maximum load that the 
fiber bundle can carry does not increase proportional
to the number of fibers, but slower. 
This is to be expected since the probability of
finding somewhere a stretch of 
weak fibers that start a fatal avalanche increases
when the number of fibers increases.

The $N/\ln N$ dependence agrees with a previous estimate by Zhang and Ding
\cite{Zhang2} and is seen also in the model with thresholds zero or unity
\cite{Duxbury,Harlow2}.
\section{Concluding remarks}
\label{sec4}
We have in this article discussed burst distributions in fiber bundles
with two different mechanisms for load distribution when fibers rupture,
viz. global or extremely local load redistributions.

The main results are the following:\\
(i) For the global model the burst distribution follows a universal power law
$\Delta^{-\frac{5}{2}}$.\\
(ii) Deviations from this power-law dependence may, however, occur for
exceptional distributions of fiber strengths.\\
(iii) For the  local model and for a uniform distribution of fiber thresholds
we show that it is possible, although complicated, to carry through an
theoretical analysis of the burst distribution.\\
(iv) A simulation study for a bundle of 20 000 fibers
confirms convincingly the theoretical results.\\
(v) For the local model the burst distribution falls off with increasing
burst size much faster than for the global model, and does not follow
a power law.\\
(vi) The expected maximum load that a bundle with global redistribution
mechanism can tolerate increases proportional to the number $N$ of fibers,
and proportional to $N/\ln N$ for the local redistribution mechanism.
\begin{appendix}
\section{}
\label{appa}
The combinatorial problem in Sec.\ \ref{sec2a} 
can be formulated more generally
as follows: Let $p_{h,n}$ be the probability that by distributing $h$
nonidentical particles among $n$ numbered boxes, box number $1$ will contain
no particles, box number $2$ will contain at most $1$ particle,and in general box
number $i$ will contain at most $i-1$ particles.

Since the probability that there 
are $h-k$ particles in box number $n$ is equal to
\[ \left( h \atop k \right) 
\left(\frac{1}{n}\right)^{h-k}\left(\frac{n-1}{n}
\right)^k, \]
we must have
\begin{equation}
 p_{h,n} = \sum_{k=0}^{h}  \left( h \atop k \right)
  \left(\frac{1}{n}\right)^{h-k}\left(\frac{n-1}{n}
\right)^k   \; p_{k,n-1}.
\label{app.2}
\end{equation}

We now prove by induction that
\begin{equation}
  p_{h,n} = 1 - \frac{h}{n}.
 \label{app.3}
 \end{equation}
Assume that this holds for 
$p_{h,n-1}$, all $h$. Insertion into the right-hand
side of (\ref{app.2})
gives
\begin{eqnarray}
 p_{h,n} &=& \sum_{k=0}^h   \left( h \atop k \right)
 \left(\frac{1}{n}\right)^{h-k}\left(\frac{n-1}{n}
\right)^k  \left(1-\frac{k}{n-1}\right)  \nonumber \\
 &=& 1 - \frac{h}{n} \sum_{k=1}^h \left( h-1 \atop k-1 \right) \left(
 \frac{1}{n}\right)^{h-k} \left(\frac{n-1}{n}\right)^{k-1} =
 1-\frac{h}{n},
 \end{eqnarray}
in accordance with (\ref{app.3}).
Since (\ref{app.3}) is valid for $n=2$, the induction is complete.

For the application in the text,
\[ p_{n-1,n} = \frac{1}{n} \]
is needed.
\section{}
\label{appb}
In Sec.\ \ref{sec3c} 
an estimate for $p(n,n,)$ was needed. We base it on the recursion
relations (\ref{n3}), (\ref{n5}) and (\ref{n7}) for $L=n+1$, $a=n$:
\begin{eqnarray}
p(n+1,n+1) &=& p_s(n+1,n)   \label{x1}\\
p_s(n+2,n+1) &=& (n+2)p_s(n+1,n)+p_2(n+2,n)  \label{x2}\\
p_2(n+3,n+1) &=& (n+1)p_s(n+1,n)+(n+1)p_2(n+2,n) \label{x3}
\end{eqnarray}
We have used that $p_s(n+1,a)$ og $p_2(n+2,a)$ vanish for $a>n$.

It is easy to eliminate $p_s$ by (\ref{x1}) and $p_2$ by (\ref{x2}), with
the result
\begin{equation}
p(n+1,n+1) = 2np(n,n)-(n-1)^2p(n-1,n-1).
\label{x4}
\end{equation}
This is a three-term recursion starting off with $p(1,1)=2$ and
$p(2,2) =2p(1,1)$ by (\ref{x4}).

With
\begin{equation}
R_n = \frac{p(n,n)}{(n-1)!}
\end{equation}
the recursion takes the form
\begin{equation}
R_{n+1}=2R_n -(1-{\textstyle \frac{1}{n}})R_{n-1}.
\label{R}
\end{equation}

Introducing the generating function
\begin{equation}
G(z) = \sum_{n=1}^{\infty} R_nz^n
\label{G}
\end{equation}
the recursion (\ref{R}) may be transformed to the differential
equation
\[ \frac{\partial}{\partial z} \left[G(z)(1-z^2)/z\right] = G(z),\]
with solution
\begin{equation}
G(z) = \frac{2z}{1-z}\;e^{z/(1-z)}.
\end{equation}
Thus the radius of convergence of the power series (\ref{G}) is unity,
and therefore
\begin{equation}
\lim_{n\rightarrow \infty} R_n^{1/n} = 1.
\label{conv}
\end{equation}

In fact $R_n \propto n^{-\frac{1}{4}}e^{2\sqrt{n}}$ for large $n$
\cite{Olaussen}.
\end{appendix}

\begin{figure}
\caption{\label{fig1} A fiber bundle with periodic boundary conditions.}
\end{figure}

\begin{figure}
\caption{\label{fig2} The threshold distribution density $p(f)$ and the 
macroscopic bundle strength $\langle F\rangle(f)$ for the distribution 
(17), with $f_0=2f_r$, and for 
$\alpha = \frac{1}{3}$ (upper curve), $\frac{1}{2}$ (middle curve), and
$\frac{2}{3}$ (lower curve).  The broken part of the $\alpha=2/3$-curve
is unstable and the macroscopic bundle strength will follow the solid
line.}
\end{figure}

\begin{figure}
\caption{\label{fig3} Burst distribution in local model as found numerically
for 4 000 000 samples with $N=20 000$ fibers ($+$), and calculated from
Eqs.\ (68) and (69) ($\circ$).  The straight line shows the power law  
$\Delta^{-5}$ and the broken curve the function $\exp(-\Delta/\Delta_0)$
with $\Delta_0=1.1$. Note the small value of $\Delta_0$.}
\end{figure}
\begin{table}
\begin{tabular}{cc}
\hline
Parameters & Asymptotics \\ 
\hline
$0 \leq f_0<1, \alpha < 1$ & 
$\Delta^{-\frac{3}{2}-\alpha} (\alpha e^{1-\alpha})^{\Delta}$ \\
$0\leq f_0 < 1, \alpha =1$ & $\Delta^{-\frac{5}{2}}$ \\
$f_0 = 1, \alpha<1$ & $\Delta^{-\frac{3}{2}}(\alpha e^{1-\alpha})^{\Delta}$\\
$1<f_0<\alpha^{-1}$ 
& $\Delta^{-\frac{5}{2}}(\alpha f_0 e^{1-\alpha f_0})^{\Delta}$ \\
$1 < f_0 =\alpha^{-1}$ & $\Delta^{-\frac{5}{2}}$ \\
$1<\alpha^{-1}<f_0$ & $\Delta^{-\frac{5}{2}} e^{-\Delta /\Delta_0}$ \\ 
\hline
\end{tabular}
\caption{\label{tab1} Asymptotic behavior of the burst distribution for 
strong threshold distributions in the global model.}
\end{table}

\begin{table}
\begin{tabular}{rrr}
\hline
$\Delta$ & Simulation & Calculation \\ 
\hline
1 & 8 327 378 752 & 8 327 331 808 \\
2 &   491 305 573 &   491 331 178 \\
3 &    72 126 803 &    72 114 644 \\
4 &    17 179 080 &    17 180 414 \\
5 &     5 590 887 &     5 591 243 \\
6 &     2 243 916 &     2 243 012 \\
7 &     1 030 833 &     1 031 678 \\
8 &       515 309 &       515 310 \\
9 &       268 589 &       268 139 \\
10 &      140 911 &       140 751 \\
11 &       72 251 &        72 701 \\
12 &       36 525 &        36 277 \\
13 &       17 523 &        17 285 \\
14 &        8 015 &         7 835 \\
15 &        3 352 &         3 392 \\
16 &        1 442 &         1 418 \\
17 &          559 &           579 \\
18 &          223 &           233 \\
19 &           90 &            93.8\\
20 &           40 &            37.5\\
21 &           18 &            15.0\\
22 &           10 &             6.0\\
23 &            1 &             2.4\\
24 &            2 &             1.0\\
25 &            0 &             0.4\\ 
\hline
\end{tabular}
\caption{\label{tab2} The burst distribution $D(\Delta)$ for the
local model with a bundle of $N=20 000$ fibers. The simulation results are
based on 4 000 000 samples.  The calculated values are based on Eqs.\ (68)
and (69), and have been multiplied by 4 000 000.} 
\end{table}
\end{document}